\documentclass[twocolumn,preprintnumbers,amsmath,amssymb]{revtex4}

\usepackage{graphicx}
\usepackage{dcolumn}
\usepackage{bm}

\newcommand{\ket}[1]{|#1\rangle}
\newcommand{\bra}[1]{\langle#1|}
\newcommand{\abs}[1]{|#1|}

\renewcommand{\i}{{\rm i}}

\newcommand{\e}{{\rm e}}

\usepackage{verbatim}
\begin{document}

\preprint{valleydonor18}


\title{Excited states of a phosphorus pair in silicon: Combining valley-orbital interaction and electron-electron interactions}
\author{W.~Wu}
\email{wei.wu@ucl.ac.uk}
\author{A. J.~Fisher}
\email{andrew.fisher@ucl.ac.uk}
\affiliation{UCL Department of Physics and Astronomy and London Centre for Nanotechnology,\\
University College London, Gower Street, London WC1E 6BT}
\date{\today}%
\begin{abstract}
Excitations of  impurity complexes in semiconductors can not only provide a route to fill the terahertz gap in optical technologies, but can also play a role in connecting local quantum bits efficiently to scale up solid-state quantum-computing devices. However, taking into account both the interactions among electrons/holes bound at the impurities, and the host band structures, is challenging. Here we combine first-principles band-structure calculations with quantum-chemistry methodology to evaluate the ground and excited states of a pair of phosphorous (shallow donors) impurities in silicon within a single framework. We account for the electron-electron interaction within a broken-symmetry Hartree-Fock approach, followed by a time-dependent Hartree-Fock method to compute the excited states. We adopt a Hamiltonian for each conduction-band valley including an anisotropic kinetic energy term, which splits the $2p_0$ and $2p_{\pm}$ transitions of isolated donors by $\sim 4$ meV, in good agreement with experiments. Our single-valley calculations show the optical response is a strong function of the optical polarisation, and suggest the use of valley polarisation to control optics and reduce oscillations in exchange interactions. When taking into account all the valleys, we have included valley-orbital interactions that split the energy levels further. We find a gap opens between the $1s\rightarrow 2p$ transition and the low-energy charge-transfer states within $1s$ manifolds (which become optically allowed because of inter-donor interactions).  In contrast to the single-valley case, we find charge-transfer excited states also in the triplet sector, thanks to the extra valley degrees of freedom.  Our computed charge-transfer excited states have a qualitatively correct energy as compared with the previous experimental findings; additionally, we predict a new set of excitations below 20 meV that have not been analysed previously. Calculations based on a statistical average of nearest-neighbour pairs at different separations suggest that THz radiation could be used to excite the donor pairs spin-selectively. Our approach can readily be extended to the other types of donors such as arsenic, and more widely to other semiconducting host materials such as germanium, zinc oxides and gallium nitride, etc.
\end{abstract}


\maketitle
\section{Introduction}\label{sec:introduction}






Donors in silicon, as building blocks for modern electronics, have recently attracted much attention as a promising candidate for developing quantum technologies \cite{zwanenburg2013}. Electrons bound to donors in silicon have shown exceptionally long spin-lattice relaxation and spin coherence times, demonstrating great potential for quantum information processing \cite{steger2012, morello2010}. Recently donor molecules (DMs) have been proposed to host electron spins as quantum bits (qubits) \cite{buch2013,weber2014, broome2018, hile2018, he2019, koch2019}, because such molecules can be used to make the spin states of different molecules distinguishable owing to hyperfine interactions. This then opens a route for addressing qubits individually \cite{buch2013,hile2018}, as in the original proposal for silicon donor quantum computation by Kane \cite{kane1998}.  A two-qubit quantum gate operation and spin readout based on silicon donors have recently been demonstrated using DMs \cite{he2019, koch2019}. In addition, exchange coupling and Pauli spin blockade have been observed between two DMs (one containing two phosphorus atoms and the other three) \cite{buch2013, weber2012, weber2014}, paving the way towards universal multi-qubit operations and qubit readout. The measurement of spin correlations and tuning of the exchange interactions between spins of different donor molecules shed light on the control of exchange interactions for two-qubit operations by using silicon donors \cite{broome2018}. However, the exchange interaction between donors in semiconductors is short-range (limited by the exponential decay of the ground states) and in many materials strongly oscillating; this is a significant obstacle to fault-tolerant quantum error correction in this system \cite{ladd2010}. Against this background, the excited states of dopants would be useful in a few respects: (i) to extend the wave functions and control the exchange interaction, thus producing longer-range coherence between the donor spins, and (ii) to connect individual qubits through an optical network by means of the optical excitations \cite{ladd2010}.

The reason for the oscillatory exchange in many host semiconductors is the interference between multiple conduction-band minima \cite{koiller2001}. An alternative approach is to use as host a material without valley degeneracy, such as ZnSe \cite{sanaka2009}, ZnO \cite{lp2019} and GaN \cite{yan2011} although at present these materials cannot match the quality of silicon crystals. On the other hand, this degree of freedom can provide opportunities, and the topic of valleytronics, in which the multiple valleys are used as an additional degree of freedom either in conventional electronics or to represent quantum information, has attracted much attention recently. The potential of valeytronics has been demonstrated by observing quantum interference between valleys in silicon \cite{salfi2014}, control of valley-polarized electrons in diamond \cite{isberg2013}, and using the valleys to control the spin properties in silicon \cite{ferdous2018}. Polarization of the valley degree of freedom is challenging for donor electrons, but could be achieved using applied strain, thus removing the exchange oscillations \cite{koiller2002}. 

The large dielectric constant ($\sim12$) for silicon and small effective mass means that the energy scale for donor electronic structure is tens of meV; this implies the natural optical couplings are in the terahertz region. Quantum cascade lasers based on sophisticated quantum-well nano-structures\cite{williams2007, dunn2020} have been used widely for THz radiation. While the terhertz radiation has been studied in isolated impurities in silicon \cite{lv2004,hubers2004,lynch2005}, recent first-principles calculations \cite{wu2018} have shown that for one-dimensional donor clusters (lines), the excitation energies can be as low as 10\,meV; they could therefore be an alternative source for Thz radiation. This will avoid using intricate fabrication techniques for quantum wells, meanwhile the frequency can be tuned by the donor densities. 

The interaction between electrons is crucial to understand the excited states of these multi-electron DMs. For example, for a pair of hydrogen atoms, stretching the bond between them will raise the energy of the so-called ionic  (or charge-transfer, CT) excited state, where two electrons sit on one atom leaving a hole on the other. This involves the competition between the on-site Coulomb repulsion (essentially the simplest form of electronic correlation) and long-range (classical) Coulomb attraction. The interplay between valley effects and electron-electron interactions is expected to bring forward new physics that is not present in the previous calculations in Ref.\cite{wu2018}, which were performed in the spherical band approximation. 
The optical properties of donor clusters were studied experimentally previously \cite{thomas1981}; this work identified a CT state located at $\sim 30$ meV. However, the experimental results were reported only down to $26$ meV with limited observations between 10 and 20 meV, which were claimed as the $1s_A \rightarrow 1s_T$ and $1s_A \rightarrow 1s_E$ transitions. The excited states of isolated donors in silicon were previously studied \cite{klymenko2017} within a tight-binding model that was based on a silicon band wave function computed by GW methods. Recently, the electron correlations were included in a single-donor multi-electron calculation within full configuration interaction \cite{tanka2018}. On the other hand, the excited states for the single-band Hubbard model have been studied in detail, which includes doublons and holons similar to ionic excited states \cite{jeckelmann1999, ah2008}. The excited states of multi-donor complexes have rarely been studied, although a configuration-interaction method was proposed to study the electronic structure of a neutral donor pair ($D_2^0$) \cite{saraiva2015}. The excited state of a donor cluster has also been studied in a three-donor complex consisting of two deep donors and one control donor, to see how the exchange interactions among them were affected by optical excitation \cite{wu2007}. In that case the excited state,  constructed in a single-valley hydrogenic model via $2s$ Whittaker function within a simple variational approach \cite{wu2008}, is delocalised over all the donors, thus affecting the sign and magnitude of exchange interaction. In addition, electron transport properties of donor arrays in silicon (involving charged, rather than neutral, excitations) have recently been studied within an extended Hubbard model \cite{le2017}.
 
Here we present a series of calculations for the excited states of a phosphorus pair in silicon. We work within effective-mass theory \cite{luttinger1955, kohn1955}, expanding the envelope functions in terms of Gaussian orbitals, while explicitly treating the interactions between electrons and preserving the multi-valley nature of the problem \cite{luttinger1955, kohn1955, ning1971}. We use the Hartree-Fock approach, and its time-dependent version, to compute the ground and excited states, respectively. We also take into account the central-cell corrections (CCC) \cite{ning1971, gamble2015, saraiva2015} to effective-mass theory, which can be adjusted according to donor types. Based on our chosen basis set, our calculations show a qualitatively correct physics, in which the nature of the lowest excitations is qualitatively different from those found in the previous hydrogenic calculations \cite{wu2018}, with a significant energy gap between the $1s\rightarrow 2p$ transition and the ionic-state transition arising from multi-valley effects both for the singlet and triplet spin sectors. In addition, we have also found that the low-energy excitation energy sector below 20 meV is dominated by the CT states, which could play an important role in exciting these donor pairs spin-selectively. The remaining discussion falls into three parts: in \S\ref{sec: computationaldetails}, we discuss the theoretical and computational methods used, in \S\ref{sec: results} we report and discuss our results, and in \S\ref{sec: conclusion}, we draw some general conclusions.

\section{Computational details}\label{sec: computationaldetails}


  


\subsection{First-principles calculation for bulk silicon}\label{sec:bandstructure}
We have performed first-principles calculations for the electronic structure of silicon by using the plane-wave code Quantum Espresso \cite{qe}. We have adopted the silicon lattice constant as $a=5.43\,\mathrm{\AA}$ with a face-centre cubic symmetry. We have chosen the GIPAW (Gauge Including Projector Augmented Waves) pseudo-potential for silicon provided in Quantum Esppresso \cite{qe}, which is compatible with the PBE exchange-correlation density functional \cite{pbe}. The Monkhorst-Pack sampling \cite{mp} of reciprocal space is carried out choosing a grid of shrinking factor equal to $16\times16\times16$. The energy cut-off is chosen to be 1088 eV. After benchmarking the silicon band structure, we have extracted the Bloch wave functions ($\phi_{\vec{k}}(\vec{r})=e^{i\vec{k}\cdot\vec{r}}\sum_{\vec{K}}c^{\vec{k}}_{\vec{K}}e^{i\vec{K}\cdot\vec{r}}$) at the conduction band minima ($|\vec{k}| = 0.85\frac{2\pi}{a}$), which were then used to compute the valley-orbital interaction. We have also performed a phase shift for the wave functions as stated in Ref.\cite{saraiwa2011} to maintain the cubic symmetry.
 
\subsection{Gaussian expansion and basis set}
Gaussian functions are used to expand the effective-mass envelope function for each valley, as in conventional molecular quantum chemistry calculations \cite{mest, mqc}. We write
\begin{eqnarray}
\Psi_\mu(\vec{r})&=&F_\mu(\vec{r})\phi_{\vec{k}_\mu}(\vec{r})\\
F_\mu(\vec{r}) &=& \sum_n c_{n\mu} g_{n\mu}(\vec{r}),
\end{eqnarray}
where $F_\mu$ is the envelope function in valley $\mu$ (the label $\mu$ runs over $\pm x$, $\pm y$ and $\pm z$), $g_{n\mu}$ is the $n$th Gaussian function for the $\mu-$valley, and $\phi_{\vec{k}_\mu}(\vec{r})=\e^{\i\vec{k}_\mu\cdot\vec{r}}u_{\vec{k}_\mu}(\vec{r})$ is the Bloch wave function for the minimum of the $\mu-$valley. We can therefore define  a state $\psi_{n\mu}$ associated with each Gaussian basis function:
\begin{eqnarray}\label{eq:psi}
\Psi_\mu(\vec{r}) &=& \sum_n c_{n\mu}\psi_{n\mu}(\vec{r})\\
\psi_{n\mu}(\vec{r}) &=& g_{n\mu}(\vec{r})\phi_{\vec{k}_\mu}(\vec{r}).
\end{eqnarray}
For the multi-valley calculations, we construct the full state from linear combinations of the single-valley states, so
\begin{equation}
\Psi(\vec{r})=\sum_\mu \Psi_\mu(\vec{r})=\sum_{\mu,n}c_{n\mu} g_{n\mu}(\vec{r})\phi_{\vec{k}_\mu}(\vec{r}).
\end{equation}

We adopt a value $11.7$ for the dielectric constant for silicon, which leads to Ha$^*= 37.77$ meV and $a_0^* = 3.26$ nm. We have chosen the typical shallow donor, phosphorus (P), throughout our calculations. For the single-valley calculations with or without central-cell correction, we use an extended even-tempered basis set (Table~\ref{tab:bs}) and benchmark our results against the electronic structure of a hydrogen atom (a single phosphorus donor in silicon) for the case without (with) central-cell correction, whereas for the multi-valley calculations we employ a moderate even-tempered basis set (Table~\ref{tab:bs}) and benchmark against the electronic structure of a single phosphorus donor in silicon. By using the single-valley basis set, we have obtained satisfactory $1s$ and $2p$ energies ($E_{1s}=-13.59$ eV, $E_{2p}=-3.40$ eV) for hydrogen atom (-18.89 meV and -4.73 meV within effective mass theory for a hydrogenic impurity). For the single-valley calculations with a central-cell correction, the ground-state energy is tuned to be -45.5 meV by varying the CCC radius as shown in Table~\ref{tab:bs}. We use a much more localized basis set in multi-valley calculations than those for the single-valley calculations. For multi-valley calculations, this basis set gives a reasonable match to the single-donor energy levels ($ E_{1sA}= -45.55$ meV, $E_{1sT}= -33.54$ meV, $E_{1sE} = -25.59$ meV, $E_{2p_0}=-8.04$ meV and $E_{2p_{x,y}}=-0.65$ meV) with our empirically chosen CCC. The basis set is designed to be moderate to have efficient multi-valley calculations; our results show that this basis set is effective. Notice that the CCC radius is chosen to fit the ground-state energy (the six $1s$-manifold ground-state energies) for single-valley (multi-valley) calculations.

\begin{table}
\centering
\begin{tabular}{llllll}
\hline
&&Single-valley&&Multi-valley&\\
Shell& BF&Exp. ($a_0^{*-2}$)&&Exp.($a_0^{*-2}$)&\\
\hline
$S$ &1& 10.0 &  & 200.0 &  \\
 &2&  3.030& &  66.667&  \\
 &3&  0.918& &  22.222& \\
 &4&  0.278&  &7.407&\\
 &5&  0.0843& &2.469&\\
 &6&  0.0256&&& \\
 &7&  0.00774&&&  \\
 &8&  0.00235& &&\\
 &9&  0.000711&&&\\
  &10&  0.000215& &&\\
 \hline
$P$ &1 & 1.0 & &20.0&\\
 &2&  0.303&&  6.667&  \\
 &3&  0.0918&  &  2.222&  \\
 &4&  0.0278&  &0.741&\\
 &5&  0.00843&  &0.247&\\
 &6&  0.00256&  &&\\
 &7&  0.000774&  &&\\
 &8&  0.000235&  &&  \\
 &9 & 0.0000711& &&\\
 &10 & 0.0000215&  &&  \\
 \hline  
CCC&$r_{cc}(a_0^*)$& 0.0199&&0.0109&\\
 \hline
\end{tabular}
\caption{The Gaussian basis sets used to perform single-valley and multi-valley calculations for the phosphorus donor. Here BF is the basis function index, and Exp is the exponent. All the contraction coefficients are $1.0$. $r_{cc}$ is the radius of the central-cell correction for the donor ion potential, defined in eq.\ref{eq:ccc} for each basis set.}\label{tab:bs}
\end{table}

\subsection{Single-valley Hamiltonian}
In contrast to the isotropic hamiltonian used in Ref.\cite{wu2018}, we explicitly include the anisotropy of the kinetic energy operator in the single-valley Hamiltonian, which therefore reads in the effective atomic units ($a_0^*$ and $\mathrm{Ha}^*$)
\begin{equation}\label{eq:mutlidonors}
\hat{H}_{u}=\sum_{i, A}[-\frac{1}{2}\nabla_i^2 + \frac{1-\gamma}{2}\frac{\partial^2}{\partial u_i^2} -\frac{1}{\abs{\vec{r}_i-\vec{R}_A}}]+\sum_{i < j}\frac{1}{\abs{\vec{r}_i-\vec{r}_j}},
\end{equation}
where $A$ runs over all the donor sites, $i$ and $j$ label electrons, and $u_i$ runs through the Cartesian coordinates $x_i$, $y_i$, and $z_i$ of each electron. $\gamma = \frac{m_\perp}{m_\parallel}$ is the ratio between perpendicular and parallel effective masses. Standard molecular \textit{ab initio} computational methods, including configuration-interaction (CI) \cite{mqc}, time-dependent Hartree-Fock (TDHF) \cite{stratmann1998} and time-dependent density-functional theory (TDDFT) \cite{stratmann1998}, can be used to compute excited states. Here we have chosen HF to compute ground states, followed by TDHF \cite{stratmann1998} for excited states. To describe the singlet ground state of a donor pair, we use the broken-symmetry method \cite{noodleman1980} to localise the spins when the donor distance becomes large. Notice that in our single-valley calculations we have neglected the interference factor from Bloch wave functions in the expansion.

\subsection{Multi-valley Hamiltonian and matrix elements}\label{sec:multivalley}
Based on the above single-valley Hamiltonian, the multi-valley hamiltonian is
\begin{equation}\label{eq:multivalley}
\hat{H}_{\mathrm{mv}}=\sum_{u}\ket{u}[\hat{H}_u+\hat{V}_{cc}]\bra{u} + \sum_{u\neq w}\ket{u}\hat{V}_{uw}\bra{w},
\end{equation}
Here $u$ and $w$ run over all the silicon conduction-band valleys $\pm x$, $\pm y$, and $\pm z$. $\hat{V}_{uw}$ includes the inter-valley interaction defined in Ref.\cite{shindo1975, gamble2015} and the contributions from electron-electron interaction ($\hat{V}_{uw}=\hat{V}_{uw}^{\mathrm{VO}}+\hat{v}^{12}_{uw}$), and in defining the operator $\hat{V}_{uw}^{\mathrm{VO}}$ we take into account only the intra-donor inter-valley splitting as in Ref.\cite{koiller2002}, (neglecting the inter-donor inter-valley interactions) . The interactions were computed in combination with first-principles calculations, from which the plane-wave coefficients of the conduction-band wave functions were extracted. The intra-donor inter-valley matrix elements were then computed as follows,
\begin{eqnarray}
\hat{V}_{uw}^{\mathrm{VO}} &&= \phi_u (\vec{r})^*U(r)\phi_w(\vec{r})\\\nonumber
&&=\sum_{\vec{K},\vec{K}^\prime} c_{\vec{K}}^wc_{\vec{K}^\prime}^{u*}U(r)e^{i[(\vec{k}_w-\vec{k}_u+\vec{K}-\vec{K}^\prime)\cdot\vec{r}]},
\end{eqnarray}
where $U(r)$ is the external potential for a single donor (with or without central-cell corrections, as discussed later) and $c_{\vec{K}}$ is the plane-wave expansion coefficient. If $U(r) = \frac{1}{r}$, then we will have a Dawson-type integral between Gaussian orbitals \cite{dawson}. For example, the matrix element between s-type Gaussian ($g_s$) orbitals reads
\begin{widetext}
\begin{equation}
\bra{g_s(\vec{r},\alpha_1)}\hat{V}_{uw}^{\mathrm{VO}} \ket{g_s(\vec{r},\alpha_2)} \\
= \sum_{\vec{K},\vec{K}^\prime} c_{\vec{K}}^wc_{\vec{K}^\prime}^{u*}N_{\alpha_1}N_{\alpha_2}\frac{4\pi F_{\mathrm{Dawson}}(\frac{\abs{\vec{k}_w-\vec{k}_u+\vec{K}-\vec{K}^\prime}}{2\sqrt{\alpha_1+\alpha_2}})}{\sqrt{\alpha_1+\alpha_2}\abs{\vec{k}_w-\vec{k}_u+\vec{K}-\vec{K}^\prime}},
\end{equation}
\end{widetext}
where $\alpha_{1,2}$ are the Gaussian exponents and $N_{\alpha_{1,2}}$ are the normalisation factors.

For the CCC, we adopt a simple Gaussian-type potential as follows, for computational convenience:
\begin{equation}\label{eq:ccc}
V_{cc}(r) = (\frac{1}{\epsilon_0}-\frac{1}{\epsilon_{\mathrm{Si}}}) e^{-r^2/r^2_{cc}}.
\end{equation}
Here $\epsilon_0$ ($\epsilon_{\mathrm{Si}}$) is the dielectric constant for the vacuum (silicon) and $r_{cc}$ is a core radius parameter that is adjusted to match the experimental binding energy for phosphorus.  This form of potential ensures that the donor electrons see a screened potential at long range, but a bare hydrogenic potential at short range; because the CCC is itself of Gaussian form, this ansatz also makes calculations of matrix elements between Gaussian basis states straightforward. The CCC formalism can be improved by changing to $e^{-\alpha r}$, but this is not the main concern of this paper.

For the Gaussian matrix elements of the overlap and one-electron ($V_1$) and two-electron repulsion ($\frac{1}{r_{12}}$) integrals, we adopt a single-valley rotating-wave approximation.
\begin{widetext}
\begin{eqnarray}\label{eq:int}
\bra{\psi_\mu(\vec{r}-\vec{R}_1)}V_1\ket{\psi_\nu(\vec{r}-\vec{R_2})}&=&v_1 e^{i\vec{k}_\mu\cdot(\vec{R}_1-\vec{R}_2)}\delta_{\mu\nu},\\
\bra{\psi_\mu(\vec{r}_1-\vec{R}_1)\psi_\nu(\vec{r}_2-\vec{R}_2)}\frac{1}{r_{12}}\ket{\psi_\gamma(\vec{r}_1-\vec{R}_3)\psi_\eta(\vec{r}_2-\vec{R}_4}&=&v_{12} e^{ik_\mu\cdot(\vec{R}_1-\vec{R}_3)+ik_\nu\cdot(\vec{R}_2-\vec{R}_4)}\delta_{\mu\gamma}\delta_{\nu\eta}.
\end{eqnarray}
\end{widetext}
Here $\psi_\mu(\vec{r})$ is as defined in eq.\ref{eq:psi}, but with the Gaussian expansion index $n$ suppressed, while $v_1$ and $v_{12}$ are the matrix elements for the one-electron and two-electron operators between Gaussian functions. As shown in the eq.\ref{eq:int}, the electron-electron interactions can contribute the inter-valley interaction as $\hat{v}^{12}_{uw}$.

All the one-electron and two-electron integrals are approximated as in Ref.\cite{koiller2002}. These integrals, arising from Gaussian functions, are computed by using Hermite integrals in a recursive manner \cite{mest}. The core Hamiltonian formed by the one-electron interactions reads
\begin{widetext}
\center
\begin{equation}
\begin{pmatrix}
H_x+V_{cc}&V^{\mathrm{VO}}_{x,-x}&V^{\mathrm{VO}}_{x,y}&V^{\mathrm{VO}}_{x,-y}&V^{\mathrm{VO}}_{x,z}&V^{\mathrm{VO}}_{x,-z}\\
V^{\mathrm{VO}}_{-x,x}&H_{-x}+V_{cc}&V^{\mathrm{VO}}_{-x,y}&V^{\mathrm{VO}}_{-x,-y}&V^{\mathrm{VO}}_{-x,z}&V^{\mathrm{VO}}_{-x,-z}\\
V^{\mathrm{VO}}_{y,x}&V^{\mathrm{VO}}_{y,-x}&H_{y}+V_{cc}&V^{\mathrm{VO}}_{y,-y}&V^{\mathrm{VO}}_{y,z}&V^{\mathrm{VO}}_{y,-z}\\
V^{\mathrm{VO}}_{-y,x}&V^{\mathrm{VO}}_{-y,-x}&V^{\mathrm{VO}}_{-y,y}&H_{-y}+V_{cc}&V^{\mathrm{VO}}_{-y,z}&V^{\mathrm{VO}}_{-y,-z}\\
V^{\mathrm{VO}}_{z,x}&V^{\mathrm{VO}}_{z,-x}&V^{\mathrm{VO}}_{z,y}&V^{\mathrm{VO}}_{z,-y}&H_{z}+V_{cc}&V^{\mathrm{VO}}_{z,-z}\\
V^{\mathrm{VO}}_{-z,x}&V^{\mathrm{VO}}_{-z,-x}&V^{\mathrm{VO}}_{-z,y}&V^{\mathrm{VO}}_{-z,-y}&V^{\mathrm{VO}}_{-z,z}&H_{-z}+V_{cc}\\
\end{pmatrix}.
\end{equation}
\end{widetext}
The matrix is formed by sub-matrices with dimension $N_g\times N_d$, where $N_g$ is the number of Gaussian function for each donor and $N_d$ is the number of donors. The diagonal term is the single-valley Hamiltonian, including the self-consistent field arising from Coulomb interactions within the HF approximation, while the off-diagonal ones are the inter-valley interaction. Notice that the Coulomb interactions will enter the Fock matrix both in the intra-valley and inter-valley terms. The dimension of the whole Hamiltonian matrix is $N_v \times N_g \times N_d$, where $N_v$ is the number of valleys (6 for Si). In the HF self-consistent-field (SCF) process, we use a simple density-matrix mixing scheme to stabilise the SCF convergence. 

\subsection{Time-dependent Hartree-Fock formalism}
The time-dependent Hartree-Fock calculations are performed following the procedure described in Ref.~\cite{stratmann1998}. We represent the interaction of the electron-hole pairs by seeking solutions of the equation
\begin{equation}\label{eq:TDHF}
\begin{bmatrix}
\bf{A} & \bf{B}\\
\bf{B}^* & \bf{A}^*
\end{bmatrix}
\begin{bmatrix}
\bf{X}\\
\bf{Y}
\end{bmatrix}
=w
\begin{bmatrix}
1& 0\\
0 & -1
\end{bmatrix}
\begin{bmatrix}
\bf{X}\\
\bf{Y}
\end{bmatrix}
\end{equation}
where
\begin{eqnarray}
A_{ai,bj} &=& \delta_{ab}\delta_{ij}(\epsilon_a-\epsilon_i) +K_{ai,bj}\\
B_{ai,bj} &=& K_{ai,jb}\\
K_{st\sigma,uv\tau} &=&(\psi^*_{s\sigma}\psi_{t\sigma}|\psi^*_{v\tau}\psi_{u\tau})\nonumber\\
&&\quad-(\psi^*_{s\sigma}\psi_{u\tau}|\psi^*_{v\tau}\psi_{t\sigma}).
\end{eqnarray}

Here $i$ and $j$ ($a$ and $b$) label the occupied (virtual) states. $s$, $t$, $\mu$, and $\nu$ ($\sigma$ and $\tau$) are used to label spatial orbitals (spins). We have used the conventional round bracket notation from quantum chemistry:
\begin{eqnarray}
&&(\psi_\alpha\psi_\beta|\psi_\gamma\psi_\delta)=\\\nonumber
&&\int d\vec{r}d\vec{r}^\prime [\psi_\alpha(\vec{r})^*\psi_\beta(\vec{r})\frac{1}{\abs{{\vec{r}-\vec{r}^\prime}}}\psi_\gamma(\vec{r}^\prime)^*\psi_\delta(\vec{r}^\prime)]
\end{eqnarray}

We therefore have 
$\bf{A} =
\begin{pmatrix}
A_{\uparrow\uparrow}&A_{\uparrow\downarrow}\\
A_{\downarrow\uparrow}&A_{\downarrow\downarrow}\\
\end{pmatrix},
$ and 
$\bf{B} =
\begin{pmatrix}
B_{\uparrow\uparrow}&B_{\uparrow\downarrow}\\
B_{\downarrow\uparrow}&B_{\downarrow\downarrow}\\
\end{pmatrix},
$
where the elements of these sub-matrices are
\begin{eqnarray}
A_{\uparrow\uparrow}: && K_{ai\uparrow, bj\uparrow} + \delta_{ab}\delta_{ij}(\epsilon_{a\uparrow}-\epsilon_{i\uparrow}) \\\nonumber
 &&  K_{ai\uparrow, bj\uparrow} = (\psi_{a\uparrow}^*\psi_{i\uparrow}|\psi_{j\uparrow}^*\psi_{b\uparrow}) - (\psi_{a\uparrow}^*\psi_{b\uparrow}|\psi_{j\uparrow}^*\psi_{i\uparrow}) \\\nonumber
A_{\uparrow\downarrow}: && K_{ai\uparrow, bj\downarrow} = (\psi_{a\uparrow}^*\psi_{i\uparrow}|\psi_{j\downarrow}^*\psi_{b\downarrow})\\\nonumber
A_{\downarrow\uparrow}: && K_{ai\downarrow, bj\uparrow} = (\psi_{a\downarrow}^*\psi_{i\downarrow}|\psi_{j\uparrow}^*\psi_{b\uparrow})\\\nonumber
A_{\downarrow\downarrow}: && K_{ai\downarrow, bj\downarrow} + \delta_{ab}\delta_{ij}(\epsilon_{a\downarrow}-\epsilon_{i\downarrow}) \\\nonumber
 &&  K_{ai\downarrow, bj\downarrow} = (\psi_{a\downarrow}^*\psi_{i\downarrow}|\psi_{j\downarrow}^*\psi_{b\downarrow}) - (\psi_{a\downarrow}^*\psi_{b\downarrow}|\psi_{j\downarrow}^*\psi_{i\downarrow}) 
\end{eqnarray} 
and
\begin{eqnarray}
B_{\uparrow\uparrow}: && K_{ai\uparrow, jb\uparrow} \\\nonumber
 &&  K_{ai\uparrow, jb\uparrow} = (\psi_{a\uparrow}^*\psi_{i\uparrow}|\psi_{b\uparrow}^*\psi_{j\uparrow}) - (\psi_{a\uparrow}^*\psi_{j\uparrow}|\psi_{b\uparrow}^*\psi_{i\uparrow}) \\\nonumber
B_{\uparrow\downarrow}: && K_{ai\uparrow, jb\downarrow} = (\psi_{a\uparrow}^*\psi_{i\uparrow}|\psi_{b\downarrow}^*\psi_{j\downarrow})\\\nonumber
B_{\downarrow\uparrow}: && K_{ai\downarrow, jb\uparrow} = (\psi_{a\downarrow}^*\psi_{i\downarrow}|\psi_{b\uparrow}^*\psi_{j\uparrow})\\\nonumber
B_{\downarrow\downarrow}: && K_{ai\downarrow, jb\downarrow} \\\nonumber
 &&  K_{ai\downarrow, jb\downarrow} = (\psi_{a\downarrow}^*\psi_{i\downarrow}|\psi_{b\downarrow}^*\psi_{j\downarrow}) - (\psi_{a\downarrow}^*\psi_{j\downarrow}|\psi_{b\downarrow}^*\psi_{i\downarrow}).
\end{eqnarray}

The oscillator strengths are computed at separations corresponding to discrete silicon lattice sites (i.e., to those donor spacings that would be allowed for active substitutional impurities in the Si lattice), and then broadened to produce the plots shown by convolving with a Lorentzian broadening of $0.1$ meV for the energy direction while the distance direction is interpolated linearly in Mathematica. We have also set the upper limit for the oscillator strength to be 0.03 (0.01) for single-valley (multi-valley) calculations in order to highlight the weak (in linear optics) but interesting low-energy transitions.

\section{Results}\label{sec: results}

\subsection{Single-valley calculations}
\subsubsection{Singlet states}\label{sec:sv-singlet}
\begin{figure}[htbp]
\includegraphics[width=10.5cm, height=7.5cm, trim={4cm 2cm 0.0cm 0.0cm},clip]{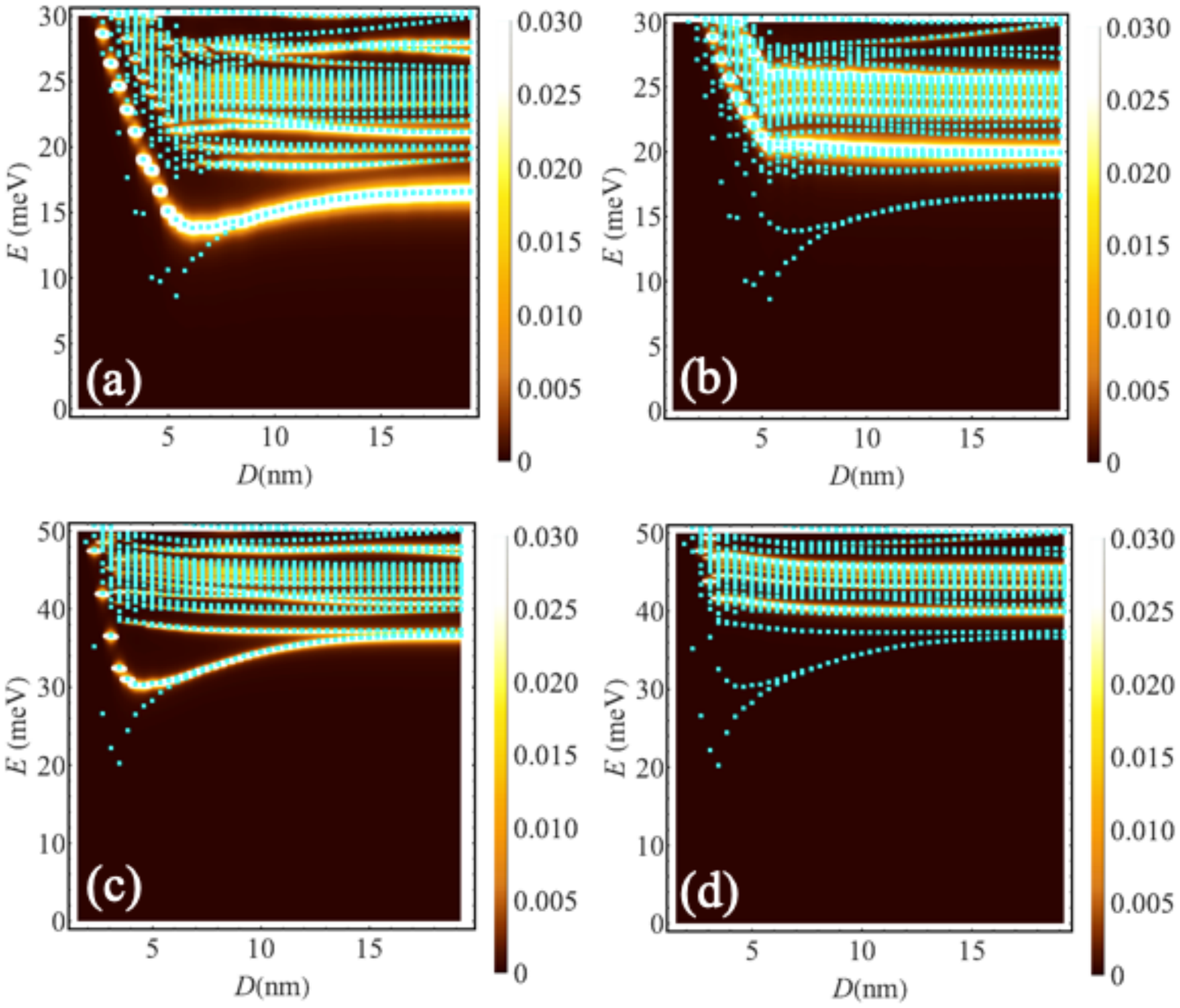}\\
\caption{(Colour online.) The singlet-state oscillator strength of a phosphorus pair in silicon along the [$\overline{1}$01] direction within a single valley is shown as a function of donor distances and excitation energies. (a) $x$-valley with $x$-polarisation of light, (b) $x$-valley with $y$-polarisation, (c) $x$-valley with $x$-polarisation with CCC, (d) $x$-valley with $y$-polarisation with CCC (Note different scale). The oscillator strength is broadened as described in the text, while the excitation energies within TDHF (solutions of equation~(\ref{eq:TDHF})) are shown as the cyan filled squares.}\label{fig:1}
\end{figure}

We have computed the ground and excited states of the single-valley Hamiltonian for a donor pair, including the excitation energies and the oscillator strengths for excitations by light with different polarisations. 
First, we exclude the central-cell potential and consider a donor pair oriented along the $[\overline{1}01]$ direction in the cubic cell ([100] in the fcc primitive cell). Here the Cartesian axes are along the three lattice vectors in the cubic cell. If we take the polarisation to lie along one of the Cartesian axes, there are five possible inequivalent combinations of the valley index and the polarisation direction; we show results for the oscillator strength in two of these cases , where the valley direction has a component along the inter-donor axis and the polarisation direction is either parallel to the valley or perpendicular to both the valley and the inter-donor axis, in Fig.\ref{fig:1} as a function of energy and donor separation.  The first case, where the valley and the polarisation are parallel ($x$-valley with $x$-polarisation or $z$-valley with $z$-polarisation) is shown in Fig.\ref{fig:1}(a). In this case, the lowest dipole-allowed excitation converges to the $1s\rightarrow 2p_0$ excitation of an isolated donor for large separations. A similar long-range limit is seen when the valley and polarization axes are parallel, but now perpendicular to the inter-donor axis (not shown).   However, when the valley and polarization axes are perpendicular ($x$-valley with $y$-polarization, Fig.\ref{fig:1}(b)), the long-range limit of the lowest allowed transition is instead the $1s\rightarrow 2p_\pm$ excitation of a single donor.  We find the splitting between the $2p_0$ and $2p_\pm$ states for an isolated P donor is $\sim 3.7$\,meV, which is in good agreement with experiment \cite{murdin2012}. (This is further supported by the calculations including CCC, Fig.\ref{fig:1}(c) and (d), discussed below).  A significant number of excitations that are not optically active can also be seen (shown as filled cyan squares in both figure panels).

The shorter-range behaviour is quite different in cases (a) and (b).  For the case where the light polarization has a component along the inter-donor axis (Fig.~\ref{fig:1}(a)), we see a characteristic branch of optically active excitations that drops down in energy below those of an isolated donor as the separation drops below approximately 6\,nm, reaching a minimum of approximately $14$ meV. We identify the transitions with minimum excitation energies as CT states, as shown in the previous work \cite{wu2018}. Their oscillator strength dominates the spectrum when the polarisation and the valley are parallel (but is much weaker when the polarisation and the valley are perpendicular).  There is no signature of the CT state in the optical response when the polarisation is perpendicular to the inter-donor axis (Fig.~\ref{fig:1}(b)), because now the light cannot couple to the CT process. 

We also show in Fig.~\ref{fig:1}(c,d) how the situation changes when the CCC is included, for the same valley and polarisation orientations.  The primary effect of the CCC is to lower the energy of the ground state while leaving the others relatively unaffected, so the main difference in the excitation spectrum is to raise all the excitation energies.  However, the dip in the CT excitation is now substantially deeper when the poliarization has a component along the dimer axis (Fig.~\ref{fig:1}(c)) and produces a minimum optically allowed excitation energy  $\sim30$ meV, approximately in agreement with the previous findings \cite{thomas1981}.  Once again, the lowest (CT) branch dominates the oscillator strength in this case, but there is no optical excitation of this branch when the polarization is perpendicular to the donor axis (Fig.~\ref{fig:1}(d)).


\subsubsection{Triplet states}

\begin{figure}[htbp]
\includegraphics[width=9cm, height=7cm, trim={2cm 0cm 0.0cm 0.0cm},clip]{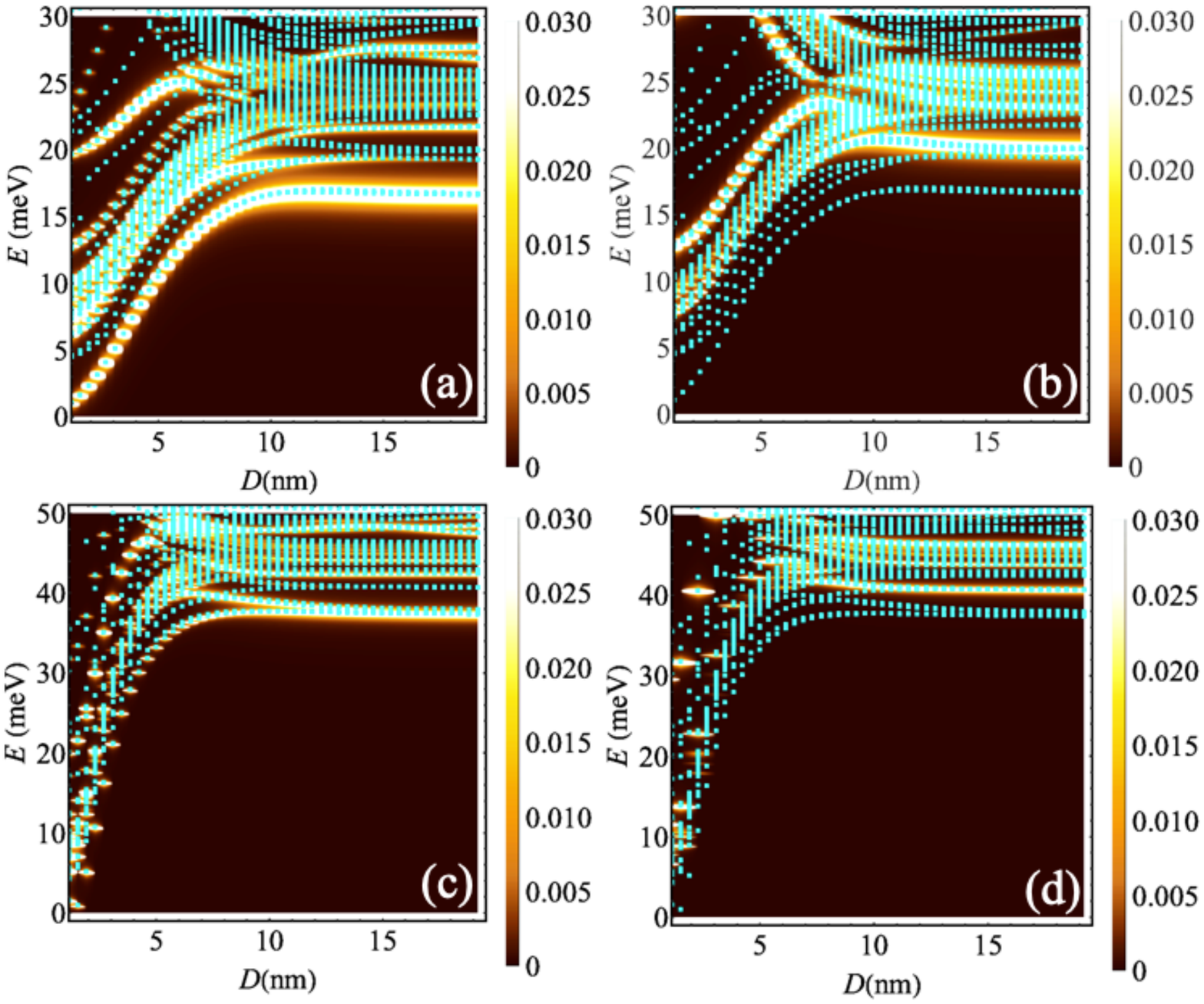}\\
\caption{(Colour online.) The triplet-state oscillator strength of a phosphorus pair within a single valley along [$\overline{1}$01] direction is shown as a function of donor distances and excitation energies. (a) $x$-valley with $x$-polarisation of light, (b) $x$-valley with $y$-polarisation, (c) $x$-valley with $x$-polarisation with CCC, (d) $x$-valley with $y$-polarisation with CCC (note different scale).  The oscillator strength is broadened as described in the text, while the excitation energies within TDHF (solutions of eq.~(\ref{eq:TDHF})) relative to the lowest triplet state are shown as the cyan filled squares.}\label{fig:2}
\end{figure}

We have also performed calculations for the triplet excitations within a single valley.  We show the results for a $[\overline{1}01]$ pair in Fig.~\ref{fig:2}, for the same combinations of valley and polarization directions as in Fig.~\ref{fig:1}. They also converge to excitations of isolated donors at large separations, but have quite different behaviour from the singlet excitations at short distances, with a collapse in the lowest excitation energy for separations below $\sim5\,\mathrm{nm}$.  A similar behaviour is observed for a pair in the hydrogenic limit \cite{wu2018} and the reason can be understood by considering the molecular orbitals of the complex: in order to form the triplet, a $1s$($\sigma^*$) anti-bonding state has to be occupied, but as the separation og the donor cores tends to zero this state becomes a $2p$ state of the He-atom analogue which has a three-fold orbital degeneracy.  This degeneracy at small separations persists even in the presence of the CCC (Fig.~\ref{fig:2}(c) and (d)), although the excitation energies are raised to $\sim 37\,\mathrm{meV}$ at long range as expected. At mid range (between 5 and 10\,nm) we find the excited states contain a mixture of $s$ and $p$ orbitals. At long inter-donor distances, we again observe the splitting between $2p_0$ and $2p_{\pm}$ transitions, with the former being excited by light polarized parallel to the valley and the latter by light perpendicular to the valley. 


\subsection{Multi-valley calculation}

\subsubsection{Singlet states}
\begin{figure}[htbp]
\includegraphics[width=9.8cm, height=4.8cm, trim={2.cm 4cm 0.0cm 4.0cm},clip]{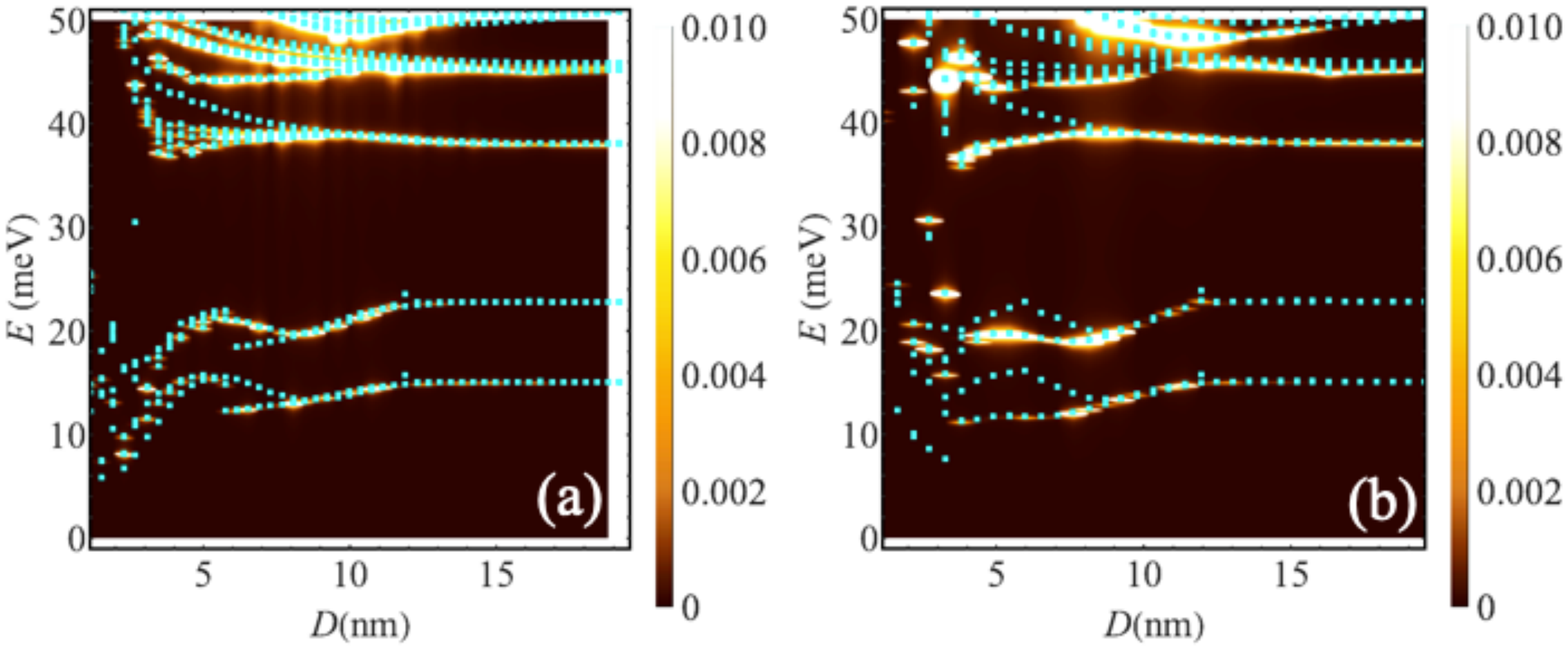}
\caption{(Colour online.) The multi-valley singlet-state oscillator strength as a function of phosphorus distance and excitation energies when donors are arranged along [$\overline{1}$01] and [100] directions. (a) [$\overline{1}$01] pair direction with $x$-polarisation of light and (b) [100] pair direction with light polarisation along donor axis.  The oscillator strength is broadened as described in the text, while the excitation energies within TDHF (solutions of equation~(\ref{eq:TDHF})) are shown as cyan filled squares.}\label{fig:3}
\end{figure}

We have also performed multi-valley calculations of a phosphorus pair for the broken-symmetry approximation to the spin-singlet state using the Hamiltonian (\ref{eq:multivalley}). For the ground-state calculations, the imbalance in spin composition of the wave function components near the two donors due to the broken symmetry states starts to emerge at separations $\sim 5$ nm and becomes dominant at $\sim 10$ nm, leading to the localisation of the opposite spins on different donors. The oscillator strengths are shown as a function of donor separation in Fig.~\ref{fig:3}(a) and (b) for the $[\overline{1}01]$ and $[100]$ pair orientations in the cubic cell, respectively. We have chosen a light polarisation along the $x$ axis (i.e. having a component along the pair axis).   Note that additional weak optical transitions appear well below the $1s \rightarrow 2p$ excitations (Fig.\ref{fig:3}(a)); further examination shows that these transitions converge to the single-donor $1s_A \rightarrow 1s_T$ and $1s_A \rightarrow 1s_E$ excitation energies in the long-range limit; for isolated donors these transitions are dipole-forbidden, but they are rendered allowed by inter-donor interactions. At shorter distances these transitions mix with a CT character; analysis of the corresponding wave functions suggests that these CT states can be derived within the 1$s$-manifold, and are formed by an electron hopping from the $1s_{A,T,E}$ state on one donor to the $1s_{A,T,E}$ on the other.  These excitation branches develop splittings at separations below $\sim$10\, nm, presumably due to bonding-antibonding splittings for both pair orientations, originating from the multi-valley effect. Further calculations with the donors separated along the cubic axis, as shown in Fig.\ref{fig:3}(b), suggest that the crossover between CT and $1s\rightarrow 2p$ transitions happens at shorter distance compared with the single-valley calculations (either with or without CCC), between $5$ and $10$ nm. There are three CT-excitation branches: one crossing the $1s\rightarrow 2p$ transition and two within the $1s$ manifold that have never been observed experimentally. We cannot exclude the possibility that there may be inaccuracies due to the relatively more localised basis set used for the multi-valley calculations. However we have tested the basis set carefully in the single-donor limit (see \S\ref{sec: computationaldetails}) and it fits all the six $1s$-manifolds within the current Gaussian approximation for the CCC; we expect the qualitative features of our findings to be robust.  The complex nature of the CT excited states at short inter-donor distance is due to the interaction between excitons in different valleys; we also note that there were some qualitative differences found in our previous work on hydrogenic impurities \cite{wu2018} between the TDHF methodology and time-dependent density functional theory (TDDFT) and full configuration interaction (FCI), with the anti-crossing between the CT and $1s\rightarrow 2p$ states not fully developed within TDHF; it is possible that similar artefacts arising from the TDHF approximation are present in these calculations.

For the $y$-polarisation (not shown here), we cannot excite the CT states since the polarization has zero component along the inter-donor axis,  just as in the single-valley calculations.  The minimum energy we find here in the multi-valley CT state at short donor distance is $\sim 8$ meV, which is well below the CT excitations identified in the previous experimental findings \cite{thomas1981}. However, there is an upper band of optically active transitions at $\sim 20$ meV, which are close to those observed previously \cite{thomas1981}.  The oscillator strengths are smaller than those found in the single-valley calculations because of the oscillating behaviour for the transition matrix elements arising from inter-valley interference, similar to the oscillation of exchange interactions for donor in silicon \cite{koiller2002}.

%



\subsubsection{Triplet state}\label{sec:multivalleytriplets}
\begin{figure}[htbp]
%
%
%
%
\includegraphics[width=9.5cm, height=4.5cm, trim={1.0cm 4cm 0.0cm 4.0cm},clip]{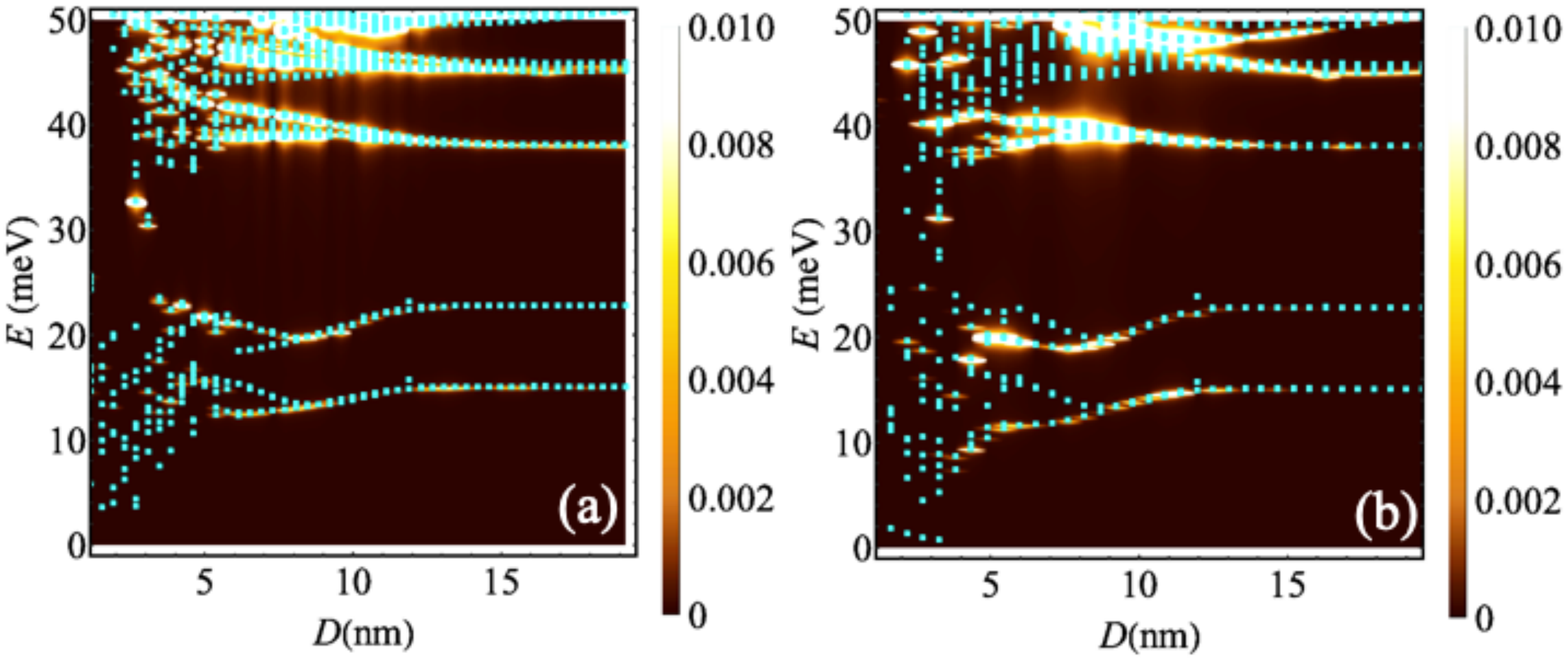}
%
\caption{(Colour online.) The multi-valley triplet-state oscillator strength as a function of donor distance and excitation energies when donors are arranged along the $[\overline{1}01]$ and [100] pair directions. (a) [$\overline{1}$01] and (b) [100] with $x$-polarisation of light.  The oscillator strength is broadened as described in the text, while the excitation energies within TDHF (solutions of equation~(\ref{eq:TDHF}) relative to the lowest triplet state are shown as cyan filled squares.}\label{fig:4}
\end{figure}

For the triplet sector (Fig.~\ref{fig:4}), we find the optical absorption now resembles that of the singlet sector more closely than was the case for single-valley calculations or the previous hydrogenic simulations \cite{wu2018}. This is because the multi-valley structure now affords more choices of low-energy states for the electrons and lessens the role of the Pauli principle in limiting the available configurations for triplets.  First, states are visible showing the characteristic distance dependence of a CT state, if the light polarisation direction is along [100] or [001] (i.e. has at least one component along the inter-donor axis). At small donor distances ($<10$ nm) we find the CT excitation energies are lower than the $1s \rightarrow 2p$ transitions; CT transitions are now allowed owing to the extra degrees of freedom provided by the valleys. The CT excitation energy is $\sim 10$ meV for an inter-donor distance of $\sim 4 $ nm, once again well below the previous experiments \cite{thomas1981}. As shown in Fig.~\ref{fig:4}, we can see a few transitions at $\sim 30$ meV, in good agreement with the previous findings \cite{thomas1981}. A confirmation of the CT nature of these triplet transitions is that $y$-polarised light (perpendicular to the axis) cannot excite them (not shown here).

A second important difference from the single-valley calculations is the lack of optically active low-energy excitations from the triplet ground state as the separation tends to zero. This is because the nature of the triplet ground state is itself different: the two electrons can now occupy different valleys, so it is no longer necessary for them to occupy an anti-bonding molecular orbital.  The other low-energy excitations have different valley structures and are 'dark', with the first optically allowed transition being to the various anti-bonding states at higher energies.  This difference is reflected in the exchange splitting between the singlet and triplet ground states, which is much lower at small separations in the multi-valley case than in the single-valley case (see \S\ref{sec:exchange}).

\subsubsection{Wave functions}

\begin{figure}[htbp]
%
%
%
%
\includegraphics[width=10cm, height=7cm, trim={5.0cm 1cm 0.0cm 1.0cm},clip]{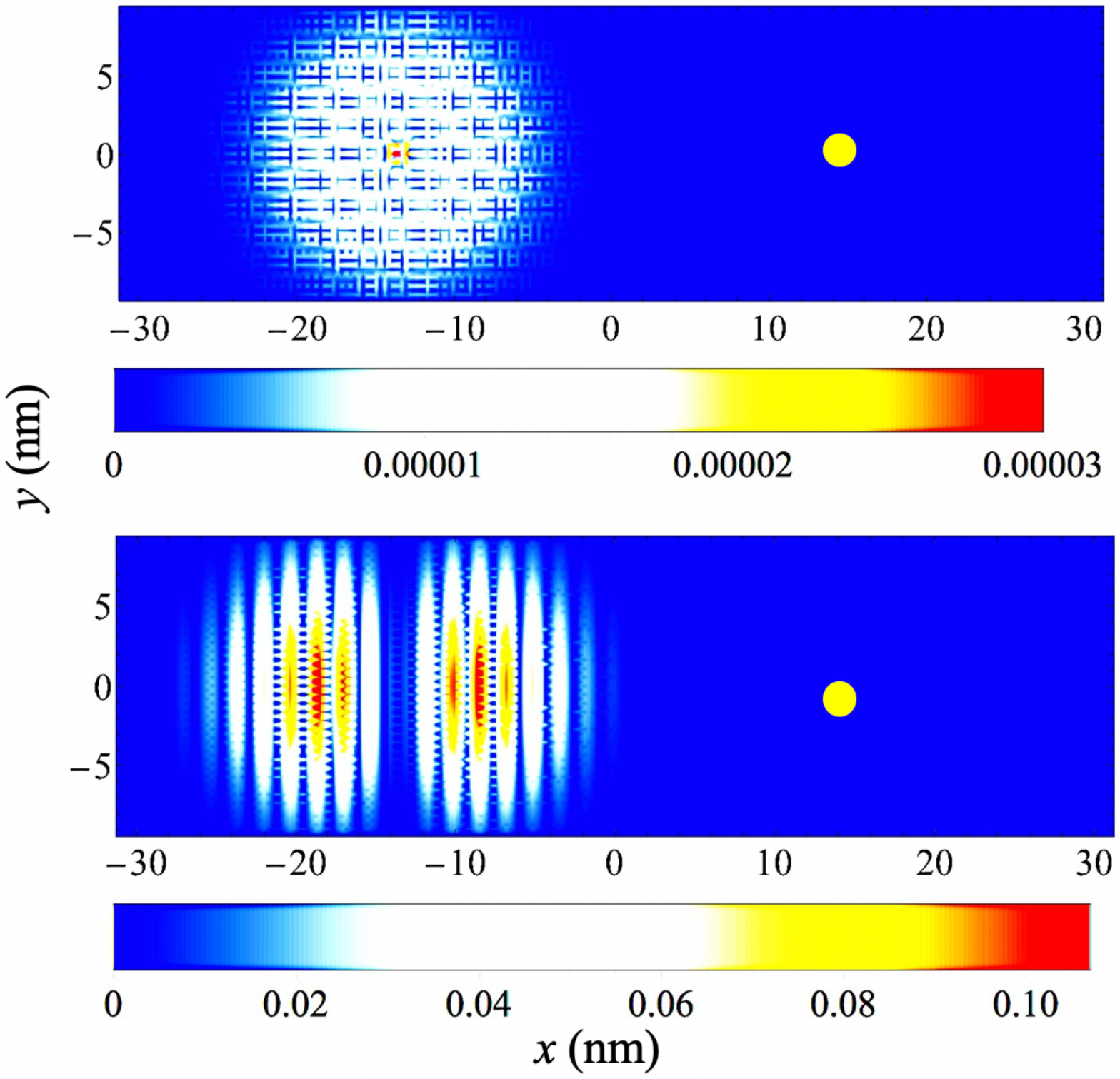}
%
\caption{(Colour online.) The HF wave functions (absolute values) at $z=0$ for the multi-valley broken-symmetry state when the inter-donor distance is large ($\sim 38.4$ nm). The upper one is the wave function for the ground state in one of the spin channels, while the lower one is the virtual $2p_x$ state in the same spin channel. Both are localised on the left donor of the pair.  The yellow dot labels the position of other (right) donor, while the colour scale displays the probability density of the state.}\label{fig:5}
\end{figure}

\begin{figure}[htbp]
%
%
%
%
\includegraphics[width=10.5cm, height=6cm, trim={4.0cm 3cm 0.0cm 3.0cm},clip]{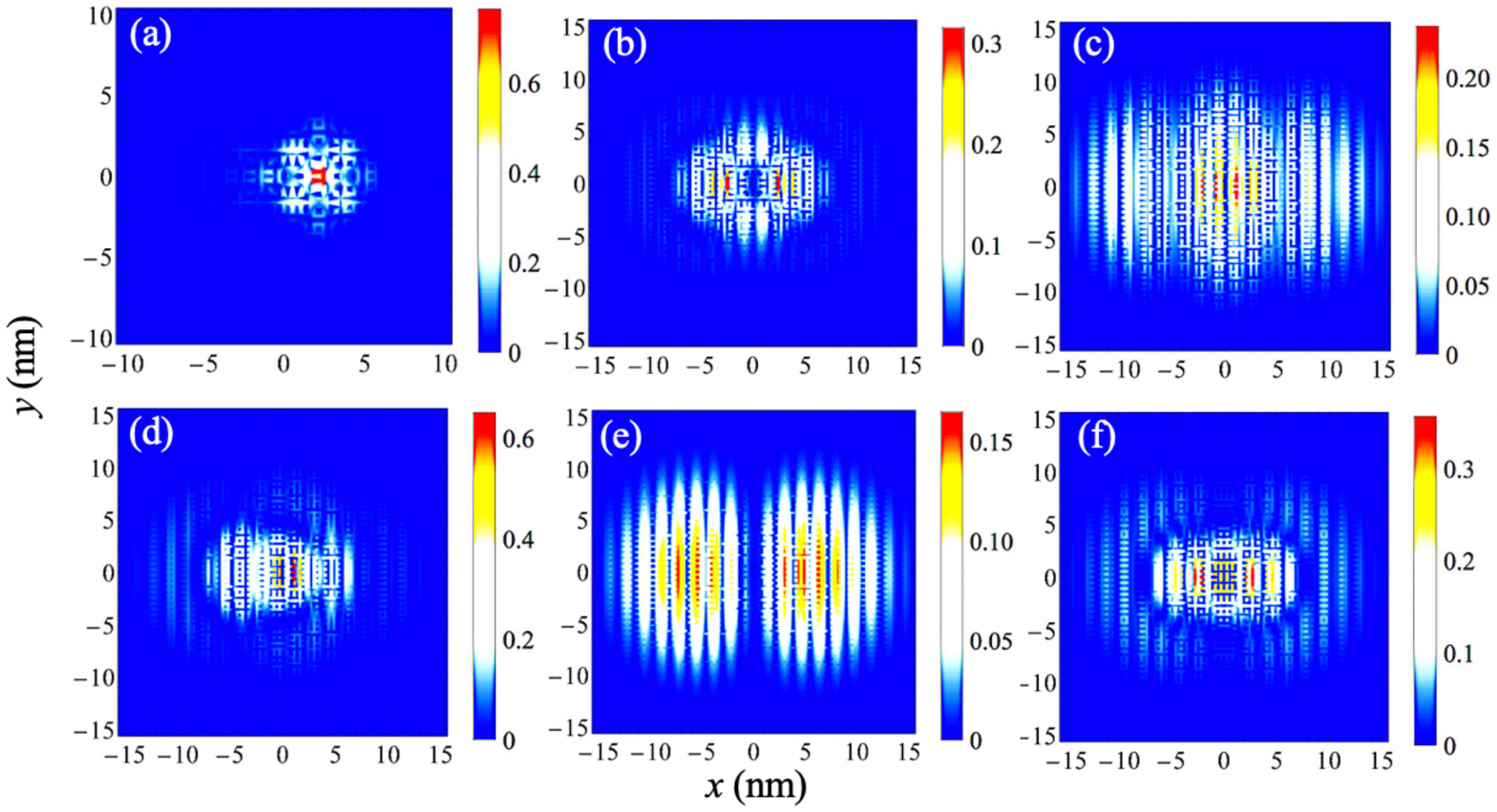}
%
\caption{(Colour online.) The ground state and dominant HF single-electron wave functions (at $z=0$) that are involved in the CT excited state, for the multi-valley singlet state when the donor distance is small ($\sim 6.1$ nm). The main features of these wave functions are derived either from $s$-orbitals or $2p_x$ orbitals, which is expected. The left donor is not shown. (a) the ground state on the donor on the right. (b)-(f) the unoccupied HF orbitals contributing to the CT excited state.}\label{fig:6}
\end{figure}

We also show the one-electron wave functions for cases with large ($\sim 38.4$ nm, Fig.~\ref{fig:5}) and small ($\sim 6.1$ nm, Fig.~\ref{fig:6}) inter-donor distances along $[\overline{1}01]$; these were chosen to show the characteristics of states in the isolated-donor limit and involved in the CT states, respectively.  We plot the absolute value of the Hartree-Fock single-electron orbitals in each case, in the $x-y$ plane cut at $z=0$.  For the large inter-donor distance shown in Fig.\ref{fig:5}, we can identify the wave functions ($1s_A$ and $2p$) for an isolated donor. For the smaller distance shown in Fig.\ref{fig:6}, we see that the unoccupied HF orbitals contributing to the excitation share the features of $1s$ or $2p_x$ orbitals, which are expected to form the main part of the CT excited state. 

We have analysed the eigenvectors of the TDHF matrices and the corresponding HF virtual orbitals involved in the CT excited states. We find that most of the dominant electron-hole pairs are formed by a localized spin on one of donors and a molecular orbital (as shown in Fig.\ref{fig:6}), which will naturally lead to a linear combination of CT and charge-resonance (CR) states. As an example, we can write down one of the electron-hole-pair components in the CT excited state for a broken-symmetry state as $c\begin{vmatrix} X_{a\uparrow}(1)& \chi^A_{i\downarrow}(1)  \\   X_{a\uparrow}(2)& \chi^A_{i\downarrow}(2) \end{vmatrix}$, where $X = \chi^A+\chi^B$ (a molecular state delocalised on both donors), $c$ is a normalisation factor, and $i$ ($a$) refers to an occupied (virtual) orbital. This determinant can then be decomposed to $[\chi^A_{a\uparrow}(1)\chi^A_{i\downarrow}(2)-\chi^A_{a\uparrow}(2)\chi^A_{i\downarrow}(1)]+[\chi^A_{a\uparrow}(1)\chi^B_{i\downarrow}(2)-\chi^A_{a\uparrow}(2)\chi^B_{i\downarrow}(1)]$. Inside the first bracket is so-called CT or ionic state, while the second one is the charge-resonance state. With the additional valley degrees of freedom, we find that the electron-hole pair can exist in different valleys, which can allow the appearance of CT excited states for the spin triplet without violating the Pauli principle. For such a triplet state, we can perform a similar wave-function analysis to obtain $[\chi^A_{a\uparrow}(1)\chi^A_{i\uparrow}(2)-\chi^A_{a\uparrow}(2)\chi^A_{i\uparrow}(1)]+[\chi^A_{a\uparrow}(1)\chi^B_{i\uparrow}(2)-\chi^A_{a\uparrow}(2)\chi^B_{i\uparrow}(1)]$, which is also a combination of CT and CR excited states. In both cases, therefore, the CT excited state is coupled to a CR excited state; the coupling strength depends on the extension of the wave function or the donor distance.

\subsubsection{Statistically averaged oscillator strength}

\begin{figure}[htbp]
%
%
%
%
\includegraphics[width=8.7cm, height=5cm, trim={0.3cm 3cm 0.0cm 2.0cm},clip]{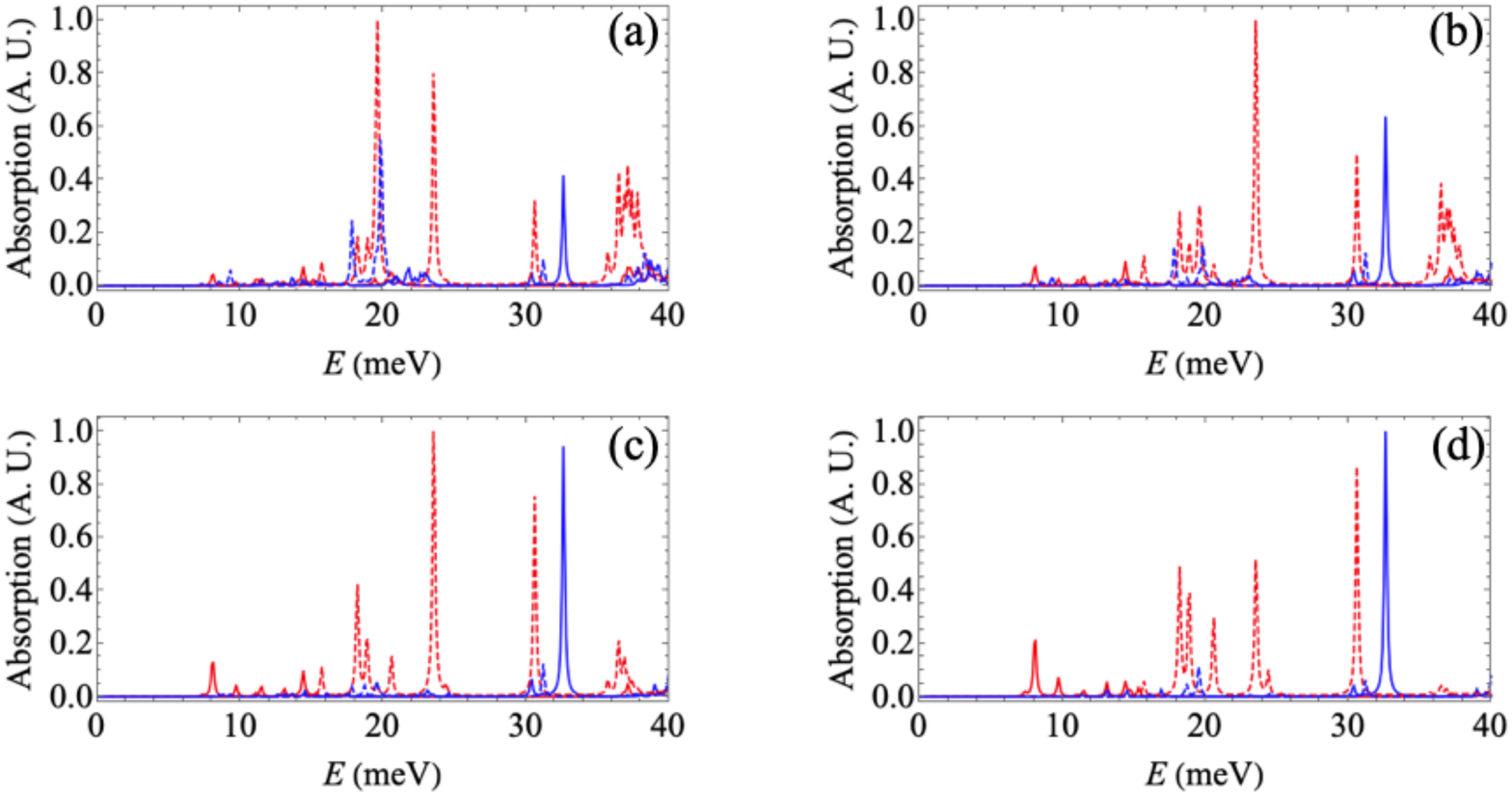}
%
\caption{(Colour online.) The normalized statistical averages of the singlet (red) and triplet (blue) pair optical spectra for different donor densities along the $[\overline{1}01]$ (solid) and the $[100]$ (dashed) directions.   The optical polarization is along the $x$-axis and the calculations are based on the multi-valley treatment of a donor pair. The chosen donor densities are (a) $5\times 10^{17}/\mathrm{cm}^3$, (b) $1\times 10^{18}/\mathrm{cm}^3$,  (c) $2\times 10^{18}/\mathrm{cm}^3$, and (d) $4\times 10^{18}/\mathrm{cm}^3$. As the densities increase, we can see (i) clearer separation of singlet and triplet excitations in a broad range of frequencies, and (ii) the emergence of low-energy CT states for both singlet and triplet sectors.}\label{fig:7}
\end{figure}

Based on the above multi-valley donor-pair calculations for both singlet and triplet states, we have performed an approximate statistical averaging of the oscillator strengths for a series of donor densities, for a range of densities where the approximation of well isolated donor pairs is valid \cite{thomas1981}.  We have used densities of $5\times 10^{17}/\mathrm{cm}^3$, $1\times 10^{18}/\mathrm{cm}^3$, $2\times 10^{18}/\mathrm{cm}^3$, and $4\times 10^{18}/\mathrm{cm}^3$, but without taking a full average over directions. To do this we have used the data presented in Figs.3--4 for individual pairs, and weighted the oscillator strengths obtained along the $[\overline{1}01]$ and $[100]$ directions with the three-dimensional nearest-neighbour distribution function for the corresponding distance from the origin \cite{sc1943, wu2018}. These calculations assume that there is no defect from any other direction having the same distance from the donor at origin. As shown in Fig.~\ref{fig:7}, for both singlet and triplet states, as the donor densities increase, the CT excitations  become more dominant over the single-donor $1s-2p$ transitions. However, the CT transitions appear in different energy ranges for the singlet and triplet: for the singlet, the dominant CT transitions are at $10-20$ meV, whereas for the triplet, they are at $\sim30$ meV.  From our results it seems likely that the experimental observations of CT transitions near 30\,meV in Ref.~\cite{thomas1981} were in fact of triplet states; we note that at the corresponding spacings of ~10\,nm, the exchange splitting is significantly smaller than $k_BT$ (of order 0.17\,meV in the experiment) and the thermal state of the pairs before excitation is therefore a classical mixture of singlets and triplets. For both directions, we find clear separation between singlet and triplet excitations over a wide range of frequencies, especially for the higher densities in Fig.~\ref{fig:7}(c) and (d): in these cases, we see particularly clear separation between singlet and triplet spectra at energies near $\sim$10, 20, and 30 meV.  This provides a broad energy window in which optical experiments such as high-resolution free-electron laser \cite{greenland2008,greenland2010} could be used to tune or interrogate the spin orientations for donor pairs.

\subsubsection{Exchange interactions}\label{sec:exchange}
\begin{figure}[htbp]
%
%
%
%
\includegraphics[width=9.3cm, height=6.5cm, trim={1.8cm 1.0cm 0.0cm 1.0cm},clip]{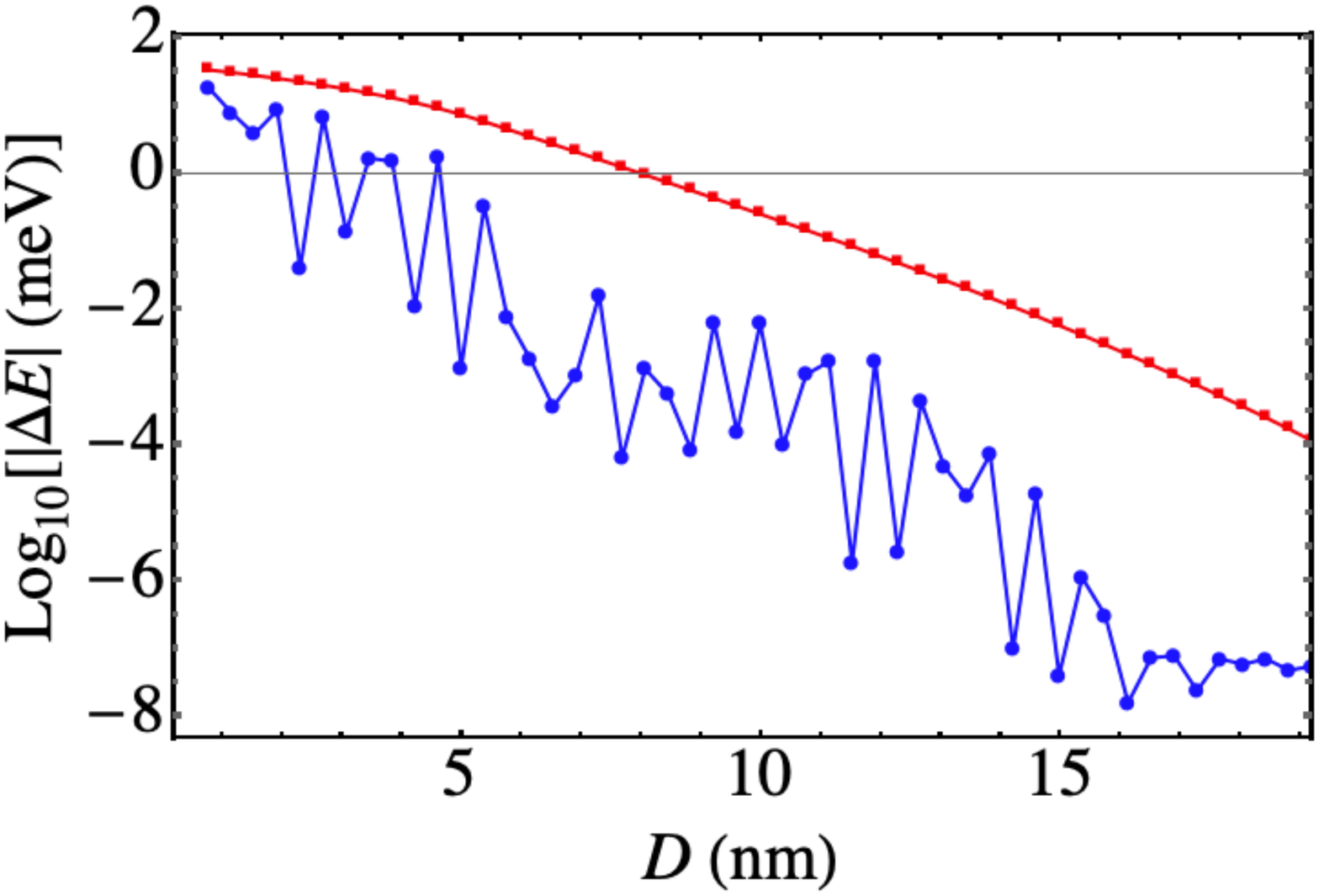}
%
\caption{ The logarithmic plot of the exchange splittings (Log$_{10}\abs{\Delta E}$) for the single-valley (x-valley, red squares) and multi-valley (blue circles) calculations along the [$\overline{1}$10] direction are shown. The exchange splitting for the multi-valley case is strongly oscillatory and much smaller than that for single valley.} \label{fig:8}
\end{figure}

We have also compared the exchange interactions for the multi-valley and single-valley ($x$-valley) cases along the [$\overline{1}$01] direction as shown in Fig.\ref{fig:8}, by directly taking energy differences between the singlet and triplet ground states. This confirms that the exchange interaction in the multi-valley case is strongly oscillatory (as previously argued on the basis of ground-state calculations \cite{koiller2001}) and shows that, even at its peak, the multi-valley exchange is much smaller than its single-valley counterpart. This illustrates the advantages of valley polarization for the suppression of exchange oscillations. At small separations this substantial difference arises because two parallel-spin electrons can occupy boding molecular orbitals in different valleys, rather than being forced to occupy an anti-bonding orbital in a single valley (see also \S\ref{sec:multivalleytriplets}).

\section{Conclusion}\label{sec: conclusion}
In this paper, we have combined first-principles band-structure calculations with quantum-chemistry methodology to compute the electronic structure, especially the excited states, of a phosphorus donor pair in a silicon-lattice environment. Within a single-valley approximation, the oscillator strengths as a function of donor distance show similar features to our previous hydrogen-cluster simulations \cite{wu2018}. From these calculations, we can also find the consistency with the experimental results for the energy gap between the $2p_{\pm}$ and $2p_0$ excited states. The single-valley calculations also show strong dependence of the optical spectra on the orientations of the valley and the polarisation vector of the light.

Our multi-valley calculations take into account the inter-valley interaction and CCC, and have been performed for several different donor axes in a silicon-lattice environment and for different light electrical-field polarisation directions. We find that both the broken-symmetry and triplet states exhibit a prominent CT state, located at an excitation energy around $\sim 30$ meV at high donor densities.  The oscillator strength in this region is dominated by triplet excitations, and the energy is approximately in agreement with the previous experimental results \cite{thomas1981}. Notice that neither our single-valley nor our multi-valley calculations shows a clear crossover of the CT states to the $D^+-D^-$ state at large separations; this is consistent with the previous results of TDHF calculations for hydrogen clusters \cite{wu2018}. There, we compared TDHF and TDDFT calculations and found that TDDFT is better for describing this crossover, possibly because of the more accurate description of electron correlations in DFT.  It is not obvious how to make a multi-valley generalisation of DFT or TDDFT; however, this finding suggests that such generalisations might be useful in the study of donor clusters. For both the broken-symmetry and triplet excited states, there are two low-energy branches of CT states converging at large separations to the energy differences between $1s_A\rightarrow 1s_E$ and $1s_A\rightarrow 1s_T$; this indicates that at intermediate distances, CT states are formed deriving entirely from the $1s$ manifold. 

As the energy scale of these excitations is close to that of exchange interactions, our calculations have pointed to using optically active CT states to control spin dynamics. Our statistical averaging calculations also show that the singlet and triplet CT excited states are relatively well separated in energy along both the lattice directions we studied; this points to the potential use of optical excitation to control, or read out, spin states of defect clusters. Compared with the previous experimental and theoretical results, our calculations shown optically active regions with CT character at substantially lower energies (typically below 20 meV), which have only been identified simply as $1s_A$ and $1s_T$ transitions \cite{thomas1981}. This shows the importance of including the valley degrees of freedom for the low-energy CT excitations; this in turn is closely related to the physics of charge transport in the donor clusters. Moreover, the algorithm and code we develop here can be readily adapted to the other defects in silicon. Looking more broadly, our calculations could be further extended to study shallow donor clusters in other semiconducting hosts with degenerate conductions band edges, such as germanium, ZnO, etc.
 
\begin{acknowledgments}
We wish to acknowledge the support of the UK Research Councils Programme under grant EP/M009564/1. We thank Jianhua Zhu, Nguyen Le, Ella Crane, Eran Ginossar, Neil Curson, and Taylor Stock, for helpful and inspiring discussions.
\end{acknowledgments}


\end{document}